\documentclass{article}
\usepackage{spconf,amsmath,amsfonts,hyperref,lipsum}
\usepackage{spconf,amsmath,graphicx,amsfonts}
\usepackage{amsmath,graphicx,amsfonts,multirow,booktabs,float,threeparttable,lipsum,dsfont,enumerate,hyperref,caption,subcaption}
\usepackage{tablefootnote}
\usepackage{blindtext}
\usepackage{hyperref}
\usepackage[table]{xcolor}

\def\s{{\mathbf{s}}}
\def\t{{\mathbf{t}}}
\def\x{{\mathbf{x}}}

\newcommand\blfootnote[1]{%
  \begingroup
  \renewcommand\thefootnote{}\footnote{#1}%
  \addtocounter{footnote}{-1}%
  \endgroup
}

\title{Identifying Source Speakers for Voice Conversion based Spoofing Attacks on Speaker Verification Systems}

\name{Danwei Cai$^1$, Zexin Cai$^1$, Ming Li$^{1,2}$}

\address{
  $^1$Department of Electrical and Computer Engineering, Duke University, Durham, USA\\
  $^2$Data Science Research Center, Duke Kunshan University, Kunshan, China\\\texttt{ming.li369@duke.edu}}

\begin{document}
\ninept

\maketitle

\begin{abstract}
An automatic speaker verification system aims to verify the speaker identity of a speech signal. However, a voice conversion system could manipulate a person's speech signal to make it sound like another speaker's voice and deceive the speaker verification system. Most countermeasures for voice conversion-based spoofing attacks are designed to discriminate bona fide speech from spoofed speech for speaker verification systems. In this paper, we investigate the problem of source speaker identification -- inferring the identity of the source speaker given the voice converted speech. To perform source speaker identification, we simply add voice-converted speech data with the label of source speaker identity to the genuine speech dataset during speaker embedding network training. Experimental results show the feasibility of source speaker identification when training and testing with converted speeches from the same voice conversion model(s). In addition, our results demonstrate that having more converted utterances from various voice conversion model for training helps improve the source speaker identification performance on converted utterances from unseen voice conversion models.
\blfootnote{Corresponding author: Ming Li}
\end{abstract}


\section{Introduction}
An automatic speaker verification system verifies the identity of speakers by analyzing audio signals.
Speaker verification is a vital biometric technology in real-world applications, such as call centers for banking systems, smartphones, smart speakers, and other Internet-of-Things devices.
As such applications are inherently security-critical, the robustness of speaker verification systems against spoofing attacks is of the utmost importance.
Even with the recent advance in speaker modeling enabled by deep neural networks (DNN) \cite{cai_exploring_2018, snyder_x-vectors:_2018}, speaker verification systems are susceptible to malicious spoofing attacks \cite{wu_spoofing_2015-1}.

Between the process of signal acquisition and verification result deliveries, the speaker verification system can be manipulated or attacked in various ways \cite{wu_spoofing_2015-1}.
Among these attacks, imposter spoofing at the microphone or during signal transmission is the most typical one.
Recent studies have demonstrated that most speaker verification systems are unprotected from various spoofing attacks, such as impersonations \cite{GONZALEZHAUTAMAKI201513}, replays \cite{villalba_detecting_2011}, speech synthesis \cite{de_leon_evaluation_2012}, voice conversion \cite{alegre_spoofing_2013}, and adversarial attacks \cite{kreuk_fooling_2018}. 
On the other hand, countermeasures have been developed to defend speaker verification systems from these spoofing attacks \cite{wu_spoofing_2015-1, li_replay_2021, todisco_constant_2017}. 
The Automatic Speaker Verification Spoofing And Countermeasures  Challenge (ASVspoof) series are being held to support independent assessments of vulnerabilities to spoofing and to assess the performance of countermeasures against spoofing \cite{wu_asvspoof_2017, yamagishi_asvspoof_2021}. 

This study focuses on the spoofing attack of voice conversion on speaker verification systems.
Voice conversion involves manipulating a speech signal of the original person (i.e., source speaker) in order to make it sound more like the speaking voice of another person (i.e., target speaker) while preserving the linguistic content \cite{mohammadi_overview_2017}.
The application of deep learning has enhanced voice conversion technologies in terms of voice naturalness and voice similarity \cite{kameoka_stargan-vc_2018-1, kaneko_cyclegan-vc2_2019}.
The improvement, however, raises concerns regarding privacy and authentication. 
Therefore, preventing the incorrect use of one's voice with voice conversion technologies becomes more and more important.

Various approaches have been proposed for speaker verification systems to defend against the spoofing attack of voice conversion. 
These countermeasures are usually designed to discriminate bona fide speech from spoofed speech for speaker verification systems.
However, none of these countermeasures provide the ability of source speaker identification -- inferring the identity of the source speaker given the voice converted speech.
Source speaker identification has potential applications in crime investigation and judicial procedures.
For example, source speaker identification can help identify a suspect involved in financial fraud with voice conversion-based impersonation spoofing. 

Typically, a voice conversion system consists of two modules -- acoustic feature conversion and waveform generation \cite{sisman_overview_2021}.
The feature conversion module changes the spectral feature of the source speaker towards the target speaker by manipulating speaker-related characteristics, such as prosody, pitch, and formants.
The waveform generation module reconstructs the time-domain speech signals from the converted spectral feature.
Generally speaking, feature conversion algorithms can be categorized into three groups - encoder-decoder models \cite{tian18_odyssey}, generative adversarial network (GAN)-based models \cite{kaneko_cyclegan-vc2_2019} and parallel spectral feature mapping models \cite{sundermann_vtln-based_2003}. 
These algorithms, however, are far from perfect.
The converted speech, to some extent, still retains the voice style of the source speaker.
For example, the speaker-related linguistic factors reflected in sentence structure, lexical choice, and idiolect are fixed and can not be altered by a voice conversion system.
For the encoder-decoder-based feature conversion, speaker-dependent factors and speaker-independent factors can not be perfectly decoupled by the encoder.
The information leak of the source speaker in the speaker-independent factors will be retained during the decoding phase.
For the GAN-based feature conversion, the use of the cycle consistency mechanism \cite{zhu2017unpaired} imposes a strict ``time-frequency bin level'' constraint during the generation process so that speaker-irrelevant features are retained \cite{li2021cvc}.

Since the voice converted speech contains information of both source and target speakers, we can train the speaker model to recognize the converted speech as the source speaker. By setting the converted speech label as the source speaker's identity, we train the speaker model with converted speech and bona fide speech, encouraging the speaker model to extract source speaker information from the converted speech while maintaining a discriminative speaker embedding space.  

From the defender's perspective, source speaker identification can be classified into white and black boxes identification. In white-box identifying, the defender can access the structure and parameters of the voice conversion models used by attackers. In black-box source speaker identification, the voice conversion models used by attackers are hidden to the defender. This paper demonstrates that source speaker identification is feasible under the white box identifying protocol: speaker verification systems can be trained to recognize the source speaker of the voice converted speech. Also, training the speaker model with converted speeches from multiple voice conversion models improves the performance of black-box source speaker identification compared to training with converted speeches generated by only one voice conversion model. This result indicates that commonality exists among converted speech from different voice conversion algorithms, and general black-box source speaker identification is possible. To the best of our knowledge, this study is the first to investigate the problem of source speaker identification for voice conversion-based spoofing attacks on speaker verification systems.

\section{Related Works}



\subsection{Voice conversion}

As mentioned before, a voice conversion system consists of two modules -- feature conversion and waveform generation \cite{sisman_overview_2021}.
The feature conversion module changes the spectral feature of the source speaker towards the target speaker; and thus takes center stage in most voice conversion studies.
Early studies on feature conversion relied on parallel speech data, i.e., the speech signals from source and target speakers share the same linguistic content.
Under the statistical modeling framework, both parametric and non-parametric approaches were applied to parallel voice conversion.
Examples of parametric approaches include Gaussian mixture model \cite{toda2007voice}, dynamic kernel partial least squares regression \cite{helander2011voice}; examples of non-parametric approaches include non-negative matrix factorization \cite{aihara2014voice} and phonetic sparse representation \cite{ccicsman2017sparse}.
However, parallel speech data is hard to collect and thus limits the application of voice conversion.
Therefore, recent studies are focused on non-parallel voice conversion.
Erro \textit{et al.} \cite{erro2009inca} proposed the INCA alignment techniques to establish correspondence between non-parallel source and target audios.
Other non-parallel feature conversion method includes the phonetic posteriograms-based approach \cite{sun2016phonetic,zheng2016text}, which uses an automatic speech recognizer to extract speaker-independent phonetic representation.
However, phonetic-based features are shown to contain speaker information in speaker verification \cite{li2016generalized}.

On the other hand, the aforementioned approaches mainly focus on one-to-one and many-to-one voice conversion, while many-to-many/any-to-any conversion greatly enhances the robustness towards unseen speakers. Various deep frameworks, including GAN \cite{gazev, stargan} and auto-encoder structures \cite{qian_autovc_2019, chen_again-vc_2021, ppp, sv}, have been employed to achieve many-to-many voice conversion. Such systems either disentangle speaker information inherently or infer discriminative speaker representation from pre-trained automatic speaker recognition systems. In this case, many-to-many voice conversion models learn the latent speaker space through seen speakers from the training data and generalize well to unseen speakers, allowing the model to clone a target voice with just one utterance from the target speaker.


\section{Methods}
Speaker embedding network aims to learn a discriminative embedding space where utterances from the same speaker are clustered together and utterances from the different speakers are separated. Voice conversion models make the attacker's speech (source speech) sound like the target speaker to deceive the speaker verification system. Therefore, converted speech is more likely to be in the subspace of the target speaker. Source speaker identification, however, aims to learn a speaker embedding space that maps converted speech to the subspace of the source speaker.

Given a source speech dataset $\mathcal{D}_s = \{\s_i\}$ and a target speech dataset $\mathcal{D}_t = \{\t_j\}$, a voice conversion model manipulate the voice of the source speech $\s_i$ to sound like the voice of target speech $\t_j$, producing a converted speech dataset $\mathcal{D}_c = \{\x_{\s_i \rightarrow \t_j } | \s_i\in\mathcal{D}_s, \t_i\in\mathcal{D}_t\}$. The source speech dataset $\mathcal{D}_s$, the target speech dataset $\mathcal{D}_t$, and the converted speech dataset $\mathcal{D}_{c,k}$ associated with voice conversion algorithm $\mathbf{C}_k$ $(k=1,\cdots,K)$ are used together as the training data $\mathcal{D}= \mathcal{D}_s \cup \mathcal{D}_t \cup \mathcal{D}_{c,1} \cup \cdots \cup \mathcal{D}_{c,K}$ to train the speaker embedding network. During training, the label of the converted speech $\x_{\s_i \rightarrow \t_j}$ is set as the speaker identity of the source speech $\s_i$. The speaker embedding network is trained to extract source speaker information from the converted speech while maintaining the discriminative speaker embedding space.

\begin{figure*}[t]
  \centering
  \includegraphics[width=\linewidth]{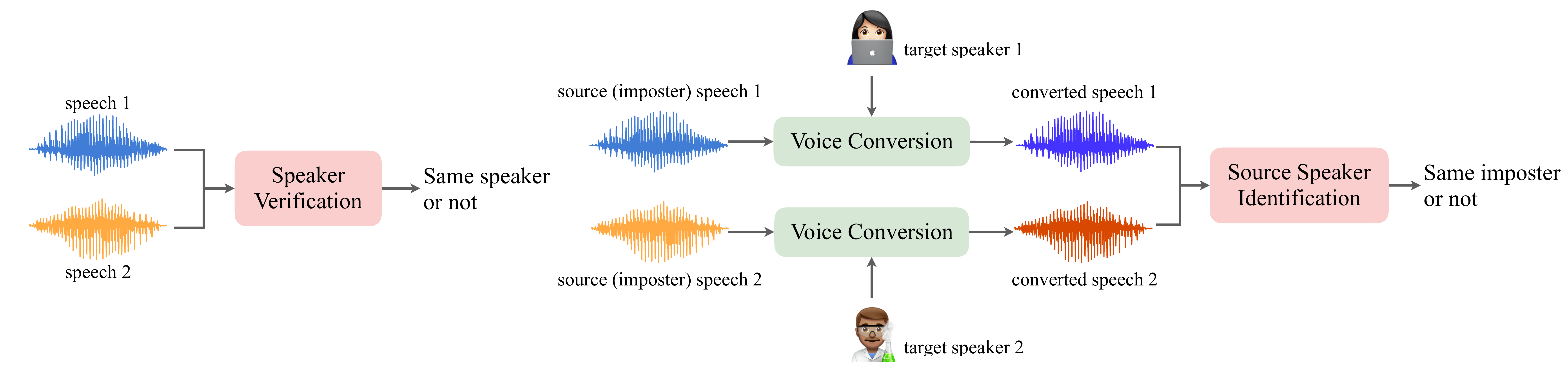}
  \caption{Speaker verification and source speaker identification.}
  \label{fig1}
\end{figure*}

\section{Experiments}

\subsection{Dataset}

We use LibriSpeech \cite{panayotov2015librispeech} as the source speech dataset and VoxCeleb \cite{nagrani_voxceleb:_2017,chung_voxceleb2:_2018} as the target speech dataset. We split LibriSpeech into a training subset containing 282,610 utterances from 2400 speakers and a testing subset containing 9,757 utterances from 84 speakers. In VoxCeleb, training data is VoxCeleb 2 development set including 1,092,009 utterances from 5,994 speakers. Testing data is VoxCeleb 1 testing set containing 4,874 utterances from 40 speakers. Training and testing data are not overlapped in both source and target datasets.

To generate voice-converted speech, we randomly sample three source speeches from LibriSpeech for one target speech in VoxCeleb. In other words, each target speech is impersonated by three different attackers. To save storage and training time, we only select one-tenth $\left(\frac{1}{10}\right)$ of VoxCeleb training data as target speech to generate converted speech for all VC systems. The total number of converted speech for a VC method is 327,600 (109,200$\times$3).

\subsection{Voice conversion models}
We use four any-to-any voice conversion algorithms, i.e., AutoVC \cite{qian_autovc_2019}, AGAIN-VC \cite{chen_again-vc_2021}, S2VC \cite{lin_s2vc_2021}, SigVC \cite{9746048}, to generate voice converted speech in the experiments.
We use the official implementation and pre-trained voice conversion model of AutoVC\footnote{\url{https://github.com/auspicious3000/autovc}}, AGAIN-VC\footnote{\url{https://github.com/KimythAnly/AGAIN-VC}}, and S2VC\footnote{\url{https://github.com/howard1337/S2VC}} to generate converted speech.
For S2VC, the self-supervised representation used in our experiments is contrastive predictive coding (CPC) \cite{oord2018representation}.
Also, instead of using the WaveRNN vocoder provided in the official implementation, we train a HiFi-GAN vocoder \cite{kong2020hifi} with VCTK dataset \cite{veaux2016superseded} for faster wave generation. 
For SigVC, we use the in-house implementation. Details of SigVC model can be found in \cite{9746048}.
All the VC systems are used to generate voice converted speech for both training and testing data.

\subsection{Speaker embedding network training}
We apply data augmentation to both genuine data (source and target speech) and voice-converted speech. Additive background noise or convolutional reverberation noise is applied to the time-domain waveform. MUSAN dataset \cite{musan} RIRS dataset \cite{ko_study_2017-1} are used for data augmentation.

During DNN training, audio waveforms in a data batch are randomly cropped between 2 to 4 seconds. Logarithmical Mel-spectrogram with 80 frequency bins is extracted as the acoustic feature. Mel-spectrograms are computed over Hamming windows of 20ms shifted by 10ms.

We use residual convolutional network (ResNet) as the speaker encoder \cite{He2016Deep}. The ResNet's output feature maps are aggregated with a global statistics pooling layer which calculates means and standard deviations for each feature map. A fully connected layer is employed afterward to extract the 128-dimensional speaker embedding. The speaker embedding network architecture is the same as in \cite{cai_within-sample_2020}. Rectified linear unit (ReLU) activation and batch normalization are applied to each convolutional layer in ResNet. Network parameters are updated using Adam optimizer \cite{kingma_adam_2017} with an initial learning rate of 0.001 along with cosine decay learning rate schedule.

\begin{table}[h]
  \caption{Speaker embedding networks trained with different training data. AutoVC, AGAIN-VC, S2VC and SigVC stand for voice converted speech generated by the corresponding algorithm. Vox refers to VoxCeleb 2 development set; Libri refers to the LibriSpeech training set.}
  \centering
  \begin{tabular}[c]{@{\ }l@{\ }c@{\ \ \ \ }c@{\ \ }c@{\ \ }c@{\ \ }c@{\ \ }c@{\ }}
  	\toprule
   	Model & Vox & Libri & AutoVC & AGAIN-VC & S2VC & SigVC\\
    \midrule
    NoVC & $\surd$ & \\
    VC1-Auto & $\surd$ & $\surd$ & $\surd$ & \\
	VC1-AGAIN & $\surd$ & $\surd$ & & $\surd$ \\
	VC1-S2 & $\surd$ & $\surd$ & & & $\surd$ \\
	VC1-Sig & $\surd$ & $\surd$ & & & & $\surd$ \\
	VC2 & $\surd$ & $\surd$ & $\surd$ & $\surd$ \\
	VC3 & $\surd$ & $\surd$ & $\surd$ & $\surd$ & $\surd$ \\
	VC4 & $\surd$ & $\surd$ & $\surd$ & $\surd$ & $\surd$ & $\surd$ \\
    \bottomrule
  \end{tabular}
  \label{tab:sys_data}
\end{table}

We train eight speaker verification systems with different setups of training data as summarized in table \ref{tab:sys_data}.
\begin{itemize}
	\item[-] System \textit{NoVC}: trained without any converted speeches
	\item[-] System \textit{VC1-$\star$}: system trained with converted speeches generated by only one voice conversion model indicated by $\star$.
	\item[-] System \textit{VC2}: trained with converted speeches generated by two voice conversion models of AutoVC and AGAIN-VC
	\item[-] System \textit{VC3}: trained with converted speeches generated by three voice conversion models of AutoVC, AGAIN-VC, and S2VC
	\item[-] System \textit{VC4}: trained with converted speeches generated by all voice conversion models
\end{itemize}
Speaker verification results and source speaker identification results are reported on equal error rate (EER).

\subsection{Source speaker identification results}

Given a pair of genuine speech, a speaker verification system verifies whether they are from the same speaker.
Given a pair of voice converted speech, source speaker identification system verifies whether they are from the same source speaker (imposter).
Figure \ref{fig1} shows the difference between speaker verification tasks and source speaker identification tasks.

For the genuine VoxCeleb 1 test set, speaker verification results are reported on the original trial list of Voxceleb 1 with 37,720 trials.
The number of true trials and false trials are the same.

For the converted speech test sets with source speaker identification tasks, a positive result is produced when the system successfully verifies the source speaker (imposter) of the converted speech.
The source speaker identification trail is based on the test trial from VoxCeleb 1.
Voice conversion is applied to convert three different source speakers into a same target voice in enrollment and test speech.
Since each target speech is impersonated by three different source speech, the total number of verification trial are 339,480 (37,720$\times$3$\times$3).
The number of true trials and false trials are 4,164 and 335,316 respectively.

\begin{table}[h]
  \caption{Verification results on EER. AutoVC, AGAIN-VC, S2VC and SigVC stand for voice converted test set generated by the corresponding algorithm. Colored cells indicate that the corresponding VC method is not used during training.}
  \centering
  \begin{tabular}[c]{@{\ \ }l@{\ \ \ \ }c@{\ \ \ \ }c@{\ \ }c@{\ \ }c@{\ \ \ \ }c@{\ \ }}
  	\toprule
   	Model & Vox1 & AutoVC & AGAIN-VC & S2VC & SigVC \\
    \midrule
    NoVC & 1.51\% & \cellcolor{yellow!30}46.1\% & \cellcolor{yellow!30}42.7\% & \cellcolor{yellow!30}45.0\% & \cellcolor{yellow!30}47.8\% \\
    VC1-Auto & 2.01\% & 15.8\% & \cellcolor{yellow!30}{40.6\%} & \cellcolor{yellow!30}{31.6\%} & \cellcolor{yellow!30}{47.6\%} \\
    VC1-AGAIN & 1.87\% & \cellcolor{yellow!30}{46.2\%} & 7.47\% & \cellcolor{yellow!30}{39.8\%} & \cellcolor{yellow!30}{46.4\%} \\
    VC1-S2 & 1.87\% & \cellcolor{yellow!30}{45.8\%} & \cellcolor{yellow!30}{36.5\%} & 6.5\% & \cellcolor{yellow!30}{46.2\%} \\
    VC1-Sig & 1.82\% & \cellcolor{yellow!30}{46.2\%} & \cellcolor{yellow!30}{40.9\%} & \cellcolor{yellow!30}{43.1\% }& 24.2\% \\
	VC2 & 2.00\% & 16.1\% & 7.83\% & \cellcolor{yellow!30}{28.0\%} & \cellcolor{yellow!30}{46.9\%} \\ 
	VC3 & 2.17\% & 16.2\% & 7.63\% & 6.37\% & \cellcolor{yellow!30}{45.0\%} \\
	VC4 & 2.15\% & 15.5\% & 8.5\% & 6.6\% & 25.0\% \\
    \bottomrule
  \end{tabular}
  \label{tab:result2}
\end{table}

We report the genuine speaker verification results and the source speaker identification results on table \ref{tab:result2}.
The first column of the table shows the results of the speaker verification on VoxCeleb 1.
We observe a performance drop of speaker verification on genuine data (VoxCeleb 1 test set) when the speaker embedding network is trained to perform source speaker identification.

From table \ref{tab:result2}, speaker embedding network trained with only genuine data (System \textit{NoVC}) can barely identify source speaker from a pair of converted speech.
Under white box testing protocol, speaker verification can be trained to recognize the source speaker in converted speech.
When training and testing with generated speeches from the same voice conversion model, system \textit{VC1-Auto}, \textit{VC1-AGAIN}, \textit{VC1-S2}, \textit{VC1-Sig} obtain EERs of 15.8\%, 7.47\%, 6.5\%, 24.2\% respectively for source speaker verification task.
It is not surprising to see a relative high EER on SigVC converted speech as SigVC system deliberately erases the speaker information from source speakers.
When training with multiple VC methods, the speaker embedding network performs source speaker identification fairly well.
For the model with all VC methods (system \textit{VC4}), source speaker verification EERs for converted speech with AutoVC, AGAIN-VC, S2VC, and SigVC are 15.1\%, 8.5\%, 6.6\%, 25.0\% respectively.

Table \ref{tab:result2} also highlights the source speaker identification results when the testing converted speech is from an unseen VC model not used during training.
Compared to system \textit{NoVC} which is trained with only genuine data, we observe performance gains on some unseen converted speech data for system \textit{VC1-Auto}, \textit{VC1-AGAIN}, \textit{VC1-S2}, and \textit{VC1-Sig}.
We can also see that system \textit{VC2} trained with AutoVC and AGAIN-VC converted speech obtains an EER of 28.0\% when performing source speaker identification on unseen S2VC converted speech data, which is 38\%, 28\%, and 10\% relative performance gain compared to system \textit{NoVC}, \textit{VC1-AGAIN}, and \textit{VC1-Auto} on unseen S2VC data, respectively.
Although this blackbox testing is far from satisfactory, it does indicate that commonality exists among converted speech from different voice conversion algorithms, and general black-box source speaker identification maybe possible with more data and advanced algorithm.

\section{Conclusion}
This study investigates the problem of source speakers identifying for voice conversion-based spoofing attacks on speaker verification systems. We train the speaker model with converted speech (labeled as the identity of source speaker) as well as genuine speech, encouraging the speaker model to extract source speaker information from the converted speech while maintaining the discriminative speaker embedding space. We demonstrate that source speaker identification is feasible when training and testing with converted speeches from the same voice conversion model. Also, when testing on converted speeches from an unseen voice conversion model, we observe that the performance of source speaker identification improves as more voice conversion models are used to generate converted speech during training. This result indicates that commonality exists among converted speech from different voice conversion algorithms, and general black-box source speaker identification is possible. Further works will investigate more voice conversion models with different model architectures and try to build a general source speaker identification system in a general black-box protocol.

\bibliographystyle{IEEEtran}
\bibliography{mybib,mybib2}

\begin{thebibliography}{10}
\providecommand{\url}[1]{#1}
\csname url@samestyle\endcsname
\providecommand{\newblock}{\relax}
\providecommand{\bibinfo}[2]{#2}
\providecommand{\BIBentrySTDinterwordspacing}{\spaceskip=0pt\relax}
\providecommand{\BIBentryALTinterwordstretchfactor}{4}
\providecommand{\BIBentryALTinterwordspacing}{\spaceskip=\fontdimen2\font plus
\BIBentryALTinterwordstretchfactor\fontdimen3\font minus
  \fontdimen4\font\relax}
\providecommand{\BIBforeignlanguage}[2]{{%
\expandafter\ifx\csname l@#1\endcsname\relax
\typeout{** WARNING: IEEEtran.bst: No hyphenation pattern has been}%
\typeout{** loaded for the language `#1'. Using the pattern for}%
\typeout{** the default language instead.}%
\else
\language=\csname l@#1\endcsname
\fi
#2}}
\providecommand{\BIBdecl}{\relax}
\BIBdecl

\bibitem{cai_exploring_2018}
W.~Cai, J.~Chen, and M.~Li, ``Exploring the {Encoding} {Layer} and {Loss}
  {Function} in {End}-to-{End} {Speaker} and {Language} {Recognition}
  {System},'' in \emph{Speaker Odyssey}, 2018, pp. 74--81.

\bibitem{snyder_x-vectors:_2018}
D.~Snyder, D.~Garcia-Romero, G.~Sell, D.~Povey, and S.~Khudanpur,
  ``x-{vectors}: {Robust} {DNN} {Embeddings} for {Speaker} {Recognition},'' in
  \emph{ICASSP}, 2018, pp. 5329--5333.

\bibitem{wu_spoofing_2015-1}
Z.~Wu, N.~Evans, T.~Kinnunen, J.~Yamagishi, F.~Alegre, and H.~Li, ``Spoofing
  and {{Countermeasures}} for {{Speaker Verification}}: A {{Survey}},''
  \emph{Speech Communication}, vol.~66, pp. 130--153, 2015.

\bibitem{GONZALEZHAUTAMAKI201513}
R.~{González Hautamäki}, T.~Kinnunen, V.~Hautamäki, and A.-M. Laukkanen,
  ``{Automatic Versus Human Speaker Verification: The Case of Voice Mimicry},''
  \emph{Speech Communication}, vol.~72, pp. 13--31, 2015.

\bibitem{villalba_detecting_2011}
J.~Villalba and E.~Lleida, ``Detecting {{Replay Attacks}} from {{Far}}-{{Field
  Recordings}} on {{Speaker Verification Systems}},'' in \emph{{European
  Workshop on Biometrics and {{ID Management}}}}, 2011, pp. 274--285.

\bibitem{de_leon_evaluation_2012}
P.~L. De~Leon, M.~Pucher, J.~Yamagishi, I.~Hernaez, and I.~Saratxaga,
  ``Evaluation of {{Speaker Verification Security}} and {{Detection}} of
  {{HMM}}-{{Based Synthetic Speech}},'' \emph{IEEE Transactions on Audio,
  Speech, and Language Processing}, vol.~20, no.~8, pp. 2280--2290, 2012.

\bibitem{alegre_spoofing_2013}
F.~Alegre, A.~Amehraye, and N.~Evans, ``Spoofing {{Countermeasures}} to
  {{Protect Automatic Speaker Verification}} from {{Voice Conversion}},'' in
  \emph{ICASSP}, 2013, pp. 3068--3072.

\bibitem{kreuk_fooling_2018}
F.~Kreuk, Y.~Adi, M.~Cisse, and J.~Keshet, ``Fooling {{End}}-{{To}}-{{End
  Speaker Verification With Adversarial Examples}},'' in \emph{{{ICASSP}}},
  2018, pp. 1962--1966.

\bibitem{li_replay_2021}
X.~Li, N.~Li, C.~Weng, X.~Liu, D.~Su, D.~Yu, and H.~Meng, ``Replay and
  {{Synthetic Speech Detection}} with {{Res2Net Architecture}},'' in
  \emph{{{ICASSP}}}, 2021, pp. 6354--6358.

\bibitem{todisco_constant_2017}
M.~Todisco, H.~Delgado, and N.~Evans, ``Constant {{Q Cepstral Coefficients}}: A
  {{Spoofing Countermeasure}} for {{Automatic Speaker Verification}},''
  \emph{Computer Speech \& Language}, vol.~45, pp. 516--535, 2017.

\bibitem{wu_asvspoof_2017}
Z.~Wu, J.~Yamagishi, T.~Kinnunen, C.~Hanil{\c c}i, M.~Sahidullah, A.~Sizov,
  N.~Evans, M.~Todisco, and H.~Delgado, ``{{ASVspoof}}: The {{Automatic Speaker
  Verification Spoofing}} and {{Countermeasures Challenge}},'' \emph{IEEE
  Journal of Selected Topics in Signal Processing}, vol.~11, no.~4, pp.
  588--604, 2017.

\bibitem{yamagishi_asvspoof_2021}
J.~Yamagishi, X.~Wang, M.~Todisco, M.~Sahidullah, J.~Patino, A.~Nautsch,
  X.~Liu, K.~A. Lee, T.~Kinnunen, N.~Evans, and H.~Delgado, ``Asvspoof 2021:
  Accelerating {{Progress}} in {{Spoofed}} and {{Deepfake Speech Detection}},''
  in \emph{{ASVspoof 2021 Workshop-Automatic Speaker Verification and Spoofing
  Coutermeasures Challenge}}, 2021.

\bibitem{mohammadi_overview_2017}
S.~H. Mohammadi and A.~Kain, ``An {{Overview}} of {{Voice Conversion
  Systems}},'' \emph{Speech Communication}, vol.~88, pp. 65--82, 2017.

\bibitem{kameoka_stargan-vc_2018-1}
H.~Kameoka, T.~Kaneko, K.~Tanaka, and N.~Hojo, ``{{StarGAN}}-{{VC}}:
  Non-parallel many-to-many {{Voice Conversion Using Star Generative
  Adversarial Networks}},'' in \emph{{{SLT}}}, 2018, pp. 266--273.

\bibitem{kaneko_cyclegan-vc2_2019}
T.~Kaneko, H.~Kameoka, K.~Tanaka, and N.~Hojo, ``Cyclegan-{{VC2}}: Improved
  {{Cyclegan}}-based {{Non}}-parallel {{Voice Conversion}},'' in
  \emph{{{ICASSP}}}, 2019, pp. 6820--6824.

\bibitem{sisman_overview_2021}
B.~Sisman, J.~Yamagishi, S.~King, and H.~Li, ``An {{Overview}} of {{Voice
  Conversion}} and {{Its Challenges}}: From {{Statistical Modeling}} to {{Deep
  Learning}},'' \emph{{IEEE/ACM Transactions on Audio, Speech, and Language
  Processing}}, vol.~29, pp. 132--157, 2020.

\bibitem{tian18_odyssey}
X.~Tian, J.~Wang, H.~Xu, E.-S. Chng, and H.~Li, ``{Average Modeling Approach to
  Voice Conversion with Non-Parallel Data },'' in \emph{{Speaker Odyssey}},
  2018, pp. 227--232.

\bibitem{sundermann_vtln-based_2003}
D.~Sundermann and H.~Ney, ``{{VTLN}}-based voice conversion,'' in \emph{{{IEEE
  workshop}} on {{Signal Processing}} and {{Information Technology}}}, 2003,
  pp. 556--559.

\bibitem{zhu2017unpaired}
J.-Y. Zhu, T.~Park, P.~Isola, and A.~A. Efros, ``Unpaired image-to-image
  translation using cycle-consistent adversarial networks,'' in \emph{ICCV},
  2017, pp. 2223--2232.

\bibitem{li2021cvc}
T.~Li, Y.~Liu, C.~Hu, and H.~Zhao, ``{CVC: Contrastive Learning for
  Non-Parallel Voice Conversion},'' in \emph{Interspeech}, 2021, pp.
  1324--1328.

\bibitem{toda2007voice}
T.~Toda, A.~W. Black, and K.~Tokuda, ``{Voice Conversion based on
  Maximum-Likelihood Estimation of Spectral Parameter Trajectory},'' \emph{IEEE
  Transactions on Audio, Speech, and Language Processing}, vol.~15, no.~8, pp.
  2222--2235, 2007.

\bibitem{helander2011voice}
E.~Helander, H.~Sil{\'e}n, T.~Virtanen, and M.~Gabbouj, ``{Voice Conversion
  using Dynamic Kernel Partial Least Squares Regression},'' \emph{IEEE
  Transactions on Audio, Speech, and Language Processing}, vol.~20, no.~3, pp.
  806--817, 2011.

\bibitem{aihara2014voice}
R.~Aihara, T.~Nakashika, T.~Takiguchi, and Y.~Ariki, ``{Voice Conversion based
  on Non-negative Matrix Factorization using Phoneme-Categorized Dictionary},''
  in \emph{ICASSP}, 2014, pp. 7894--7898.

\bibitem{ccicsman2017sparse}
B.~{\c{C}}i{\c{s}}man, H.~Li, and K.~C. Tan, ``{Sparse Representation of
  Phonetic Features for Voice Conversion with and without Parallel Data},'' in
  \emph{ASRU}, 2017, pp. 677--684.

\bibitem{erro2009inca}
D.~Erro, A.~Moreno, and A.~Bonafonte, ``{INCA Algorithm for Training Voice
  Conversion Systems from Nonparallel Corpora},'' \emph{IEEE Transactions on
  Audio, Speech, and Language Processing}, vol.~18, no.~5, pp. 944--953, 2009.

\bibitem{sun2016phonetic}
L.~Sun, K.~Li, H.~Wang, S.~Kang, and H.~Meng, ``{Phonetic Posteriorgrams for
  Many-to-One Voice Conversion without Parallel Data Training},'' in
  \emph{ICME}, 2016.

\bibitem{zheng2016text}
H.~Zheng, W.~Cai, T.~Zhou, S.~Zhang, and M.~Li, ``{Text-Independent Voice
  Conversion Using Deep Neural Network based Phonetic Level Features},'' in
  \emph{ICPR}, 2016, pp. 2872--2877.

\bibitem{li2016generalized}
M.~Li, L.~Liu, W.~Cai, and W.~Liu, ``{Generalized i-vector Representation with
  Phonetic Tokenizations and Tandem Features for both Text Independent and Text
  Dependent Speaker Verification},'' \emph{Journal of Signal Processing
  Systems}, vol.~82, no.~2, pp. 207--215, 2016.

\bibitem{gazev}
Z.~Zhang, B.~He, and Z.~Zhang, ``{GAZEV: GAN-Based Zero-Shot Voice Conversion
  Over Non-Parallel Speech Corpus},'' in \emph{Interspeech}, 2020, pp.
  791--795.

\bibitem{stargan}
R.~Wang, Y.~Ding, L.~Li, and C.~Fan, ``{One-Shot Voice Conversion Using
  Star-Gan},'' in \emph{ICASSP}, 2020, pp. 7729--7733.

\bibitem{qian_autovc_2019}
K.~Qian, Y.~Zhang, S.~Chang, X.~Yang, and M.~{Hasegawa-Johnson}, ``{{AUTOVC}}:
  {{Zero-Shot Voice Style Transfer}} with {{Only Autoencoder Loss}},'' in
  \emph{{{ICML}}}, 2019, pp. 5210--5219.

\bibitem{chen_again-vc_2021}
Y.-H. Chen, D.-Y. Wu, T.-H. Wu, and H.-y. Lee, ``{{AGAIN-VC}}: {{A One-shot
  Voice Conversion}} using {{Activation Guidance}} and {{Adaptive Instance
  Normalization}},'' in \emph{{{ICASSP}}}, 2021, pp. 5954--5958.

\bibitem{ppp}
S.~H. Mohammadi and T.~Kim, ``{One-Shot Voice Conversion with Disentangled
  Representations by Leveraging Phonetic Posteriorgrams},'' in
  \emph{Interspeech}, 2019, pp. 704--708.

\bibitem{sv}
C.~Deng, Y.~Chen, and H.~Deng, ``{One-Shot Voice Conversion Algorithm Based on
  Representations Separation},'' \emph{IEEE Access}, vol.~8, pp.
  196\,578--196\,586, 2020.

\bibitem{panayotov2015librispeech}
V.~Panayotov, G.~Chen, D.~Povey, and S.~Khudanpur, ``{Librispeech: an ASR
  Corpus based on Public Domain Audio Books},'' in \emph{ICASSP}, 2015, pp.
  5206--5210.

\bibitem{nagrani_voxceleb:_2017}
A.~Nagrani, J.~S. Chung, and A.~Zisserman, ``Voxceleb: {A} {Large}-{Scale}
  {Speaker} {Identification} {Dataset},'' in \emph{Interspeech}, 2017, pp.
  2616--2620.

\bibitem{chung_voxceleb2:_2018}
J.~S. Chung, A.~Nagrani, and A.~Zisserman, ``Voxceleb2: {{Deep Speaker
  Recognition}},'' in \emph{Interspeech}, 2018, pp. 1086--1090.

\bibitem{lin_s2vc_2021}
J.-h. Lin, Y.~Y. Lin, C.-M. Chien, and H.-y. Lee, ``{{S2VC}}: {{A Framework}}
  for {{Any-to-Any Voice Conversion}} with {{Self-Supervised Pretrained
  Representations}},'' in \emph{Interspeech}, 2021, pp. 836--840.

\bibitem{9746048}
H.~Zhang, Z.~Cai, X.~Qin, and M.~Li, ``{SIG-VC: A Speaker Information Guided
  Zero-Shot Voice Conversion System for Both Human Beings and Machines},'' in
  \emph{ICASSP}, 2022, pp. 6567--65\,571.

\bibitem{oord2018representation}
A.~v.~d. Oord, Y.~Li, and O.~Vinyals, ``Representation learning with
  contrastive predictive coding,'' \emph{arXiv preprint arXiv:1807.03748},
  2018.

\bibitem{kong2020hifi}
J.~Kong, J.~Kim, and J.~Bae, ``{HiFi-GAN: Generative adversarial networks for
  efficient and high fidelity speech synthesis},'' in \emph{NeurIPS}, vol.~33,
  2020, pp. 17\,022--17\,033.

\bibitem{veaux2016superseded}
C.~Veaux, J.~Yamagishi, K.~MacDonald \emph{et~al.}, ``{Superseded-CSTR VCTK
  Corpus: English Multi-speaker Corpus for CSTR Voice Cloning Toolkit},'' 2016.

\bibitem{musan}
D.~Snyder, G.~Chen, and D.~Povey, ``{MUSAN}: {A} {Music}, {Speech}, and {Noise}
  {Corpus},'' \emph{arXiv:1510.08484}, 2015.

\bibitem{ko_study_2017-1}
T.~Ko, V.~Peddinti, D.~Povey, M.~L. Seltzer, and S.~Khudanpur, ``{A Study on
  Data Augmentation of Reverberant Speech for Robust Speech Recognition},'' in
  \emph{{{ICASSP}}}, 2017, pp. 5220--5224.

\bibitem{He2016Deep}
K.~He, X.~Zhang, S.~Ren, and J.~Sun, ``{Deep Residual Learning for Image
  Recognition},'' in \emph{CVPR}, 2016, pp. 770--778.

\bibitem{cai_within-sample_2020}
D.~Cai, W.~Cai, and M.~Li, ``Within-{Sample} {Variability}-{Invariant} {Loss}
  for {Robust} {Speaker} {Recognition} {Under} {Noisy} {Environments},'' in
  \emph{{ICASSP}}, 2020, pp. 6469--6473.

\bibitem{kingma_adam_2017}
D.~P. Kingma and J.~Ba, ``Adam: {{A Method}} for {{Stochastic Optimization}},''
  in \emph{{ICLR}}, 2015.

\end{thebibliography}

\end{document}